# Overspend? Late? Failure?
# What the Data Say About IT Project Risk in the Public Sector

By
Alexander Budzier and Bent Flyvbjerg
BT Centre for Major Programme Management
Saïd Business School
University of Oxford

# Abstract

Implementing large-scale information and communication technology (IT) projects carries large risks and easily might disrupt operations, waste taxpayers' money, and create negative publicity. Because of the high risks it is important that government leaders manage the attendant risks. We analysed a sample of 1,355 public sector IT projects. The sample included large-scale projects, on average the actual expenditure was $130 million and the average duration was 35 months.

Our findings showed that the typical project had no cost overruns and took on average 24% longer than initially expected. However, comparing the risk distribution with the normative model of a thin-tailed distribution, projects' actual costs should fall within -30% and +25% of the budget in nearly 99 out of 100 projects. The data showed, however, that a staggering 18% of all projects are outliers with cost overruns >25%. Tests showed that the risk of outliers is even higher for standard software (24%) as well as in certain project types, e.g., data management (41%), office management (23%), eGovernment (21%) and management information systems (20%). Analysis showed also that projects duration adds risk: every additional year of project duration increases the average cost risk by 4.2 percentage points.

Lastly, we suggest four solutions that public sector organization can take: (1) benchmark your organization to know where you are, (2) de-bias your IT project decision-making, (3) reduce the complexities of your IT projects, and (4) develop Masterbuilders to learn from the best in the field.

*"For the past 40 years, for example, we've tortured ourselves over our inability to finish a software project on time and on budget. But this never should have been the supreme goal. The more important goal is transformation, creating software that changes the world or that transforms a company or how it does business… Software development is and always will be somewhat experimental."* - DeMarco, 1995[i]

**Introduction**

Government leaders know that every large IT project carries some level of risk. Many high-profile public sector projects have cost too much, taken too long, or failed to deliver the promised benefits. The consequences can include disruptions to daily operations, financial losses, negative publicity, or all of the above. Because the failure of an IT change initiative often has wide-ranging repercussions, government leaders—not just chief information officers—are beginning to understand they must proactively manage the attendant risks.

In May 1996 the Benefits Agency of the UK Department of Social Security and the executives at Post Office Counters Ltd. jointly made a decision. They awarded a one billion GBP, seven-year contract that would change how 20,000 post offices would work, and how 17 million benefits recipients would receive their money (760 million payments every year). However, the decision quickly turned sour - in July the first issues were discovered two month before the system development was planned to start, by the end of the year the project was re-baselined to allow increase the time for system development from seven to twelve months. In the end the project was abandoned in May 1999, software development was not finished and the most current forecast projected the project to deliver three years behind scheduled go-live, 30% budget overrun and a total spend of 1 billion pounds. (National Audit Office[ii])

This was an eye-catching example of failure. Whenever a large IT system goes live the press is rife with stories about it. Yet, the public sector is not alone. Our research shows that IT project failure can bring down entire private companies. But how risky are IT Projects really? Several attempts have been made to turn the anecdotal evidence of cost and schedule overruns and benefits shortfalls in large-scale IT projects into surveys that measure risk more systematically.

When, in 1999, the Standish Group first published the now infamous Chaos Report their findings seemed to epitomise the perceived IT project crisis. Academic research by Jones[iii], who aimed to develop measurements to quantify the scope of IT projects reported a plummeting success rate of such projects with increasing size. Both studies saw virtually no success for projects larger than 6 million dollars. However, academia debunked these findings. Surveys and case studies of IT portfolios failed to find signs of an IT project crisis. Many of our colleagues have questioned the rigour and criticized the opaque data collection behind the alarmist studies. Recently, our colleagues[iv] showed that using the Standish Groups definitions of 'successful', 'challenged', and 'failed projects' actually harms an organization.

Moreover, none of the prior studies has recognised the special context of the public sector (for a survey of studies see our 2011 working paper *Double Whammy*). The research project first tries to answer the question: "How risky are public sector IT projects really?"

**A global stock-take**

To help public sector leaders better understand and manage risk in IT projects, we have conducted the largest-ever global study of public sector IT change initiatives. Our research, covering a sample of 1,355 IT projects was supported by McKinsey & Company. Our data came from three sources: Freedom of Information Act requests to 26 U.S. federal departments, reports from the U.S. Government Accountability Office and international equivalents, and project documents from public sector organizations. The IT projects in our sample had an average initial budget of $130 million and an average duration of 35 months.

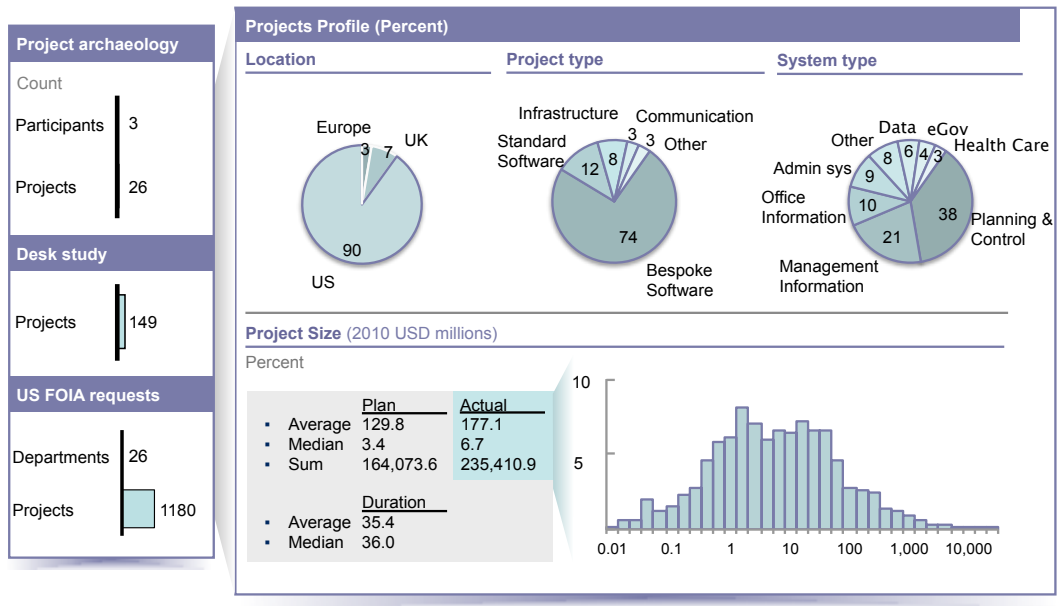

**We currently have 1,355 public sector projects in our database**

### Results and analysis

Our research is still ongoing, but our preliminary analysis has already yielded two surprising findings. The first is that public sector IT projects, on average, do not go over budget at all. This counterintuitive finding makes sense in light of our second surprising finding: that the risk distribution of IT projects is full of outliers in very fat tails and far from a normal distribution. Most risk-management models assume a normal distribution with thin tails—in this case, a "normal" project's actual costs would be within -30 percent and +25 percent of the budget (the light-gray area in Exhibit 2), and 99.3 percent of projects would fall within this range, i.e. less than 0.7 percent of all projects are outliers. Public sector IT, however, is a different story: the data show that a staggering 18 percent of all projects are "expensive" outliers (that is, projects with cost overruns above 25 percent).

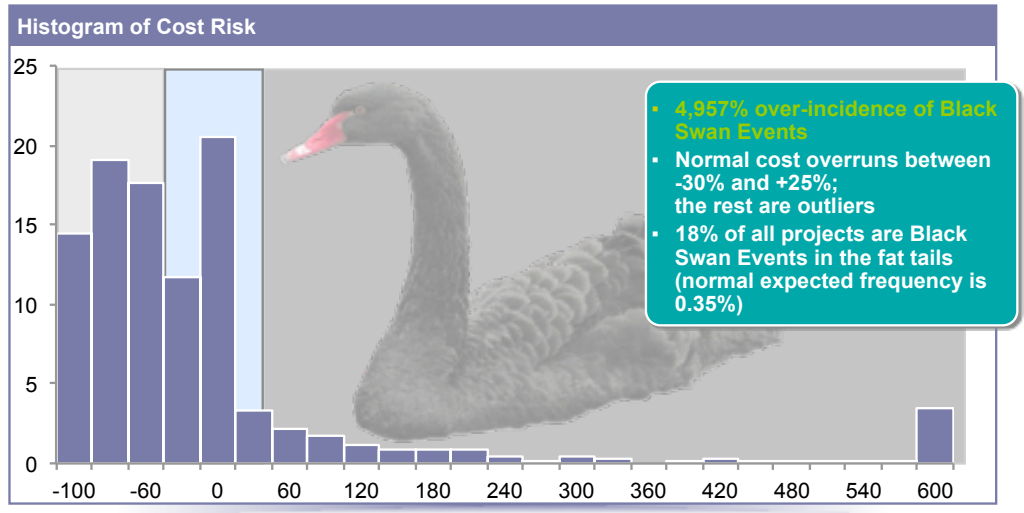

We call these expensive outliers "black swans," after Nassim Nicholas Taleb in his 2007 book "The Black Swan." Taleb defines black swans as rare, high-impact events that seem improbable and unforeseeable but, in hindsight, are explainable. In our sample we found a 4,957 percent over-incidence of black swans.

From this perspective starting an IT project is similar to playing Russian Roulette. From 6 projects proposals that ask for funding on your desk, 1 is going to run out of control so badly that it might not only bring government services to a halt, but can cause major economic damage, and end your career.

**Black swans have high cost, schedule and benefit risks, which are hidden in the fat tails of ICT portfolios**

University of Oxford
BT Centre for Major Programme Management
Saïd Business School

Risk comparison, Median, in percent

|  | Black Swans [1] | Projects with cost overrun | Normal projects | Starved projects |
|---|---|---|---|---|
| Cost overrun | +130 | +47 | +0 | -75 |
| Schedule overrun | +41 | +38 | +24 | n/a |
| Likelihood | 18 | 28 | 31 | 51 |

- Even if the median cost overrun is low, risk of cost overruns is high
- Black Swans mean very high cost and schedule risks
- And all the projects with a downside risk show significant risk

[1] The statistical expectation value does not converge and is infinite

This disproportionate number of black swans represents very high risks for public sector organizations. The typical IT project in our sample had 0 percent cost overruns and took an unremarkable 24 percent longer to implement than initially expected. When you green-light a project today the likelihood of this happening are 31%. These are not the jaw-dropping, triple-digit overruns often associated with large government IT projects. There is, however, a 28% chance that a project overruns the budget, then a reasonable expectation is that the project needs 47% more budget and 38% more time than initially expected.

But what really matters are the Black Swans. The data shows that with a probability of 18% the project is going to spin out of control and a typical Black Swan overruns its cost by 130 percent in real terms and falls 41 percent behind schedule.

Our research also uncovered another class of black swans, albeit a very different species. We found that in the last four years half of all public sector IT projects has suffered an average budget cut of 75 percent. These "starved projects," shown in the light-gray area in Exhibit 2, are also a public management challenge.

Public sector leaders must prepare robust answers to the question, "What will we do if half our ongoing IT projects get their budgets cut by as much as 75 percent?"

When it comes to assessing risk in public sector IT projects, therefore, simple averages don't matter much. What matters are the black swans—the one in six projects that could go terribly wrong. In light of the enormous risks, managing black swans can only be the responsibility of top management, not just IT leadership. A crucial question for top management becomes, "What will our organization do in the event that every sixth large IT project we undertake goes over budget by 130 percent and takes 40 percent longer than scheduled?"

Apart from bracing for impact, what can a public sector organization do?  We believe it is 4 things that count
1. Benchmark your organization to know where you are
2. De-bias your IT project decision-making
3. Reduce the complexities of your IT projects
4. Develop Masterbuilders – learn from the best in the field

First, we find that for most organizations the starting point and the most valuable knowledge comes from simply understanding where they stand. Very often, we find, organisations lack a reliable view of how the organisation's projects perform. A very first step is to establish a system of reporting that enables impactful decision-making: spot things early, surface issues quickly, act on them decisively Moreover, a powerful benchmark compares not only your project risks (incl. the risk of projects turning into a Black Swan) internally but also with other public and private sector organizations. Exhibit 4 shows a starting point how decision-makers can break down and better understand the risk faced by their organization. Our data shows, for example, that standard software projects, despite their perceived ease of implementation face the highest risk of turning into Black Swans. The data also shows that Big Data and office automation projects are particularly risky. How does your organization's risk compare?

**Despite perceived ease of implementation standard software projects have highest risk of turning into a Black Swan**

University of Oxford
BT Centre for Major Programme Management
Saïd Business School

Percent

**Black Swan Risk**

| Category | Percent |
|---|---|
| Standard software | 26% |
| Bespoke software | 18% |
| IT Architecture | 16% |
| IT Infrastructure | 14% |

18%

**Black Swans**

| Category | Percent |
|---|---|
| Data (e.g. DWH, EDI) | 41% |
| Office mgmt. (e.g. DCM, WfM, CMS) | 23% |
| eGovernment | 21% |
| MIS (e.g. GIS, CRM, MIS) | 20% |
| Administration (e.g. case/ claims mgmt.) | 17% |
| ERP (e.g. budget, facilities mgmt.) | 16% |

18%

Secondly, once you have understood and measured the risks in your organization you can turn this knowledge into better decisions. The Nobel laureate Daniel Kahneman and his colleague Amos Tversky discovered that decisions can be de-biased by taking the outside view. Simply by asking the question: How did the last 10 IT projects do that tried to do the same as the planned one? What does this say about the likelihood of success or failure for the planned project?

Also ask whether there were tell-tale signs that either a project was running smoothly or got into trouble. Our data shows, for instance, that projects that are described as unique by people involved in them have a 3-times higher propensity of turning into Black Swans. If project uniqueness is a common claim in your organization, be aware!

Thirdly, to mitigate the risks that Black Swans pose, public sector organizations must reduce the complexity of their IT projects. Our data shows quite clearly that the longer the project the higher the risk (Exhibit 5). Every additional year of project duration increases the average cost risk by 4.2 percentage points. Long projects are also more likely to turn into a Black Swan. Reducing the project complexity, for instance by introducing modularity, helps improving IT project risks.

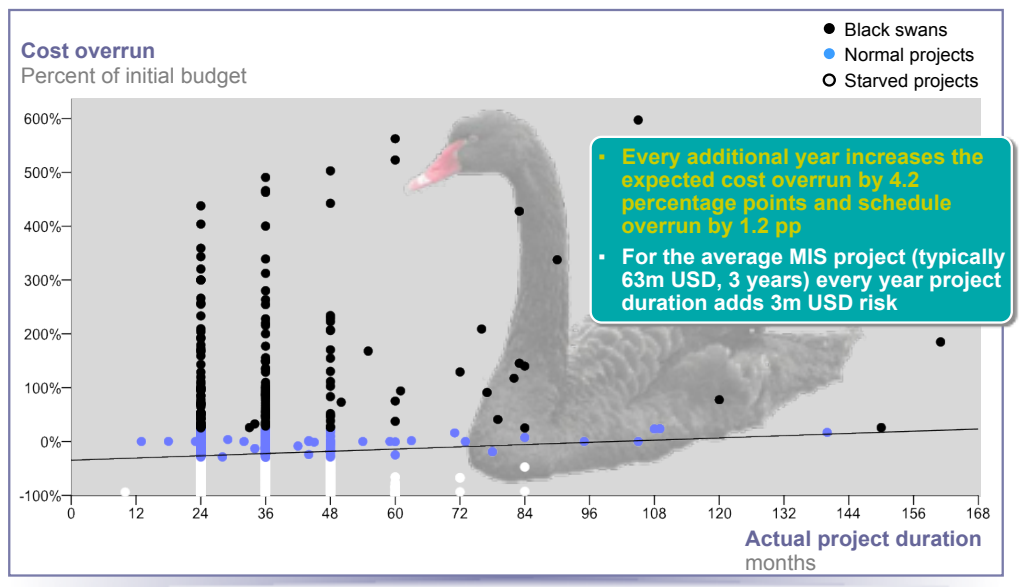

Lastly, learn from the best in the field. Learn from, what we call Master Builders. Our observations show that in every organization and in every industry there are project managers who are able to deliver on-time and on-budget time and time again. Finding your Masterbuilders, uncovering their practices of how to avoid Black Swans might seem the most difficult task but it will be the best defence of your organization and your career against a disastrous Black Swan.

[1633 words]

*Alexander Budzier* is a doctoral candidate at the BT Centre for Major Programme Management at Oxford University's Saïd Business School, and a former consultant with McKinsey's Office of Business Technology. **Bent Flyvbjerg** is the first BT Professor and founding Chair of Major Programme Management at Oxford; he is also founding Director of the BT Centre for Major Programme Management.

[i] DeMarco, T., 1995. Why does software cost so much? And other puzzles of the Information Age, New York, NY: Dorset House Publ.

[ii] National Audit Office, 2008. Department for Work and Pensions: Information Technology Programmes National Audit Office, ed. London.

[iii] Jones, C., 2008. Applied Software Measurement: Global Analysis of Productivity and Quality 3rd ed. McGraw-Hill.

[iv] Eveleens, J. & Verhoef, C., 2010. The rise and fall of the Chaos report figures. *IEEE Software*, 27(1), pp.30–36.